\documentclass[12pt]{iopart}

\usepackage{iopams}
\usepackage{graphicx}
\usepackage[dvipsnames]{xcolor}
\usepackage[normalem]{ulem}

\expandafter\let\csname equation*\endcsname\relax

\expandafter\let\csname endequation*\endcsname\relax

\usepackage{amsmath}
\usepackage[colorlinks]{hyperref}

\renewcommand{\Im}{\mathrm{Im}}

\newcommand{\eps}{\varepsilon}
\newcommand{\h}{\hbar}
\newcommand{\bm}{\boldsymbol}

\begin{document}

\title[Fermi polaron fine structure in strained van der Waals heterostructures]{Fermi polaron fine structure in strained van der Waals heterostructures}

\author{Z.A. Iakovlev \& M.M. Glazov}

\address{Ioffe Institute, 194021, St. Petersburg, Russia}
\ead{iakovlev.zakhar@gmail.com}
\vspace{10pt}

\begin{abstract}
The fine structure of attractive Fermi polarons in van der Waals heterostructures based on monolayer transition metal dichalcogenides in the presence of elastic strain is studied theoretically. The charged excitons (trions), three particle bound states of two electrons and hole or two holes and electron, do not show any strain-induced fine structure splitting compared to neutral excitons whose radiative doublet is split by the strain into linearly polarized components.  The correlation of the trions with Fermi sea holes gives rise to the attractive Fermi polarons. We show that it results in the fine structure splitting of the polaron into states polarized along the main axes of the strain tensor. The effect is related to the bosonic statistics of Fermi polarons. We develop microscopic theory of the effect and calculate the strain-induced splitting of Fermi polarons both for tungsten- and molybdenum-based monolayers identifying the role of inter- and intravalley exciton-electron interactions. The fine structure splitting of attractive Fermi polaron is proportional both to excitonic splitting and the Fermi energy. The Fermi polaron fine structure in bilayers is briefly analyzed and the role of electron and trion localization in moir\'e potentials is discussed.
\end{abstract}

%
%
%
%
%

\section{Introduction}

Van der Waals heterostructures based on atomically thin transition metal dichalcogenides show unique optical properties controlled by Coulomb-correlated electrons and holes~\cite{Geim:2013aa,RevModPhys.90.021001,Tartakovskii:2020aa}. Three-fold rotational symmetry combined with the broken inversion symmetry in monolayers results in chiral selection rules for optical transitions that enable both spin and valley orientation of charge carriers and excitons by circularly polarized light and valley entanglement by linearly polarized light~\cite{Mak:2012qf,Kioseoglou,PhysRevLett.112.047401,PhysRevB.92.075409,PhysRevLett.117.187401,Glazov_2021}. Additional degrees of freedom are provided by bilayer structures where the material composition, atomic registry, twist angle, and moir\'e potentials enable further control of optical, spin, and valley effects in these extremely two-dimensional semiconductors~\cite{Xiao_2020,Rivera:2018aa,Yu_2018,Forg:2019aa,Alexeev:2019aa,Jin:2019aa,Seyler:2019aa,Tran:2019aa,PhysRevB.105.L241406}. 

Doping of two-dimensional semiconductors with electrons or holes gives rise to novel quasiparticles, positively or negatively charged trions: Three particle bound states of two holes and electron, $X^+$, or two electron and a hole, $X^-$~\cite{Mak:2013lh,Courtade:2017a}. These quasiparticles predicted back in 1950s~\cite{PhysRevLett.1.450} have almost negligible binding energies in bulk semiconductors. They have been first observed in quasi-two-dimensional semiconductor quantum wells~\cite{PhysRevLett.71.1752,PhysRevLett.74.976} where the binding energy of excitons and trions increases due to the quantum confinement effect~\cite{Stebe:1989aa,ivchenko05a,Semina_2022}. In heterostructures based on transition metal dichalcogenides the trion binding energies are in the range of 20\ldots 40~meV~\cite{Mak:2013lh,Courtade:2017a,PhysRevB.88.045318,Kezerashvili:2016aa,2053-1583-4-2-022004,PhysRevLett.114.107401,PhysRevB.95.081301,Semina:2019aa} due to the strictly two-dimensional confinement and enhanced Coulomb interaction. It makes possible to study excited states of trions~\cite{PhysRevLett.123.167401,2020arXiv200503722J,doi:10.1063/5.0013092,PhysRevLett.125.267401} and trions formed from electrons in high-lying, excited, bands~\cite{Lin:2022aa}.

The fine structure of excitons and trions is known to be drastically different~\cite{PhysRevLett.71.1752,PhysRevB.53.R1709,astakhov99,bayer2002,PhysRevLett.109.157403,Poltavtsev:2019aa}. In contrast to neutral excitons, trions are expected to demonstrate neither fine structure splitting into linearly polarized components in the presence of the system anisotropy nor optical alignment that is linear polarization of emission induced by linearly polarized excitation. It is related to the \emph{fermionic} nature of trions that are composed of three fermions, opposed to bosonic statistics for excitons as discussed below in Sec.~\ref{sec:symmetry}. On the other hand, strictly speaking, optical excitation of a structure with a gas of charge carriers, electrons or holes, results in formation of \emph{bosonic} quasiparticles -- Fermi polarons -- where the trion is correlated with the hole in a Fermi sea~\cite{PSSB:PSSB343,Rapaport:2001uq,suris:correlation,Sidler:2016aa,PhysRevB.95.035417,PhysRevB.102.085304}. In many cases, it is a matter of convenience which approach should be used to analyze the response~\cite{Glazov:2020wf}. However, in terms of the fine structure, trions and Fermi polarons should behave drastically different, but these effects have not been studied neither theoretically, nor experimentally in sufficient detail.

In this paper we develop the theory of the energy spectrum fine structure of Fermi polarons in two-dimensional semiconductors in the presence of anisotropic strain that splits radiative doublet into linearly polarized components.  Elastic strains can be easily applied to two-dimensional semiconductors and their effects on excitonic properties are widely studied nowadays~\cite{Castellanos-Gomez:2013tn,Zhu:2013ve,2053-1583-3-2-021011,PhysRevB.98.115308,Niehues:2018tm,Chakraborty2020,Florian:ub,PhysRevB.106.125303,Hernandez-Lopez:2022aa} with wide prospects in quantum technologies. In heterostructures strains are inevitable and contribute to potential landscapes experienced by excitons and trions~\cite{Bai:2020uh,kogl2022moire}. We demonstrate that, by contrast to trions, Fermi polarons inherit linear polarization splitting from excitons. The fine structure splitting of the attractive Fermi polaron, the quasiparticle stemming from the bound trion state, is proportional to both the exciton anisotropic splitting and resident charge carriers density. Our model can be applied to mono- and bilayers of transition metal dichalcogenides as well as to the conventional semiconductor quantum well structures.

The rest of the paper is organized as follows: Sec.~\ref{sec:symmetry} presents symmetry analysis which illustrates qualitative difference between the trions and Fermi polarons in the presence of strain. Sec.~\ref{sec:model} contains the analytical model of the effect. Numerical and analytical results and their analysis are summarized in Sec.~\ref{sec:results}. Section~\ref{sec:concl} briefly summarizes key results of our work.

\section{Symmetry analysis}\label{sec:symmetry}

\begin{figure}[ht]
    \centering
\includegraphics[width=\linewidth]{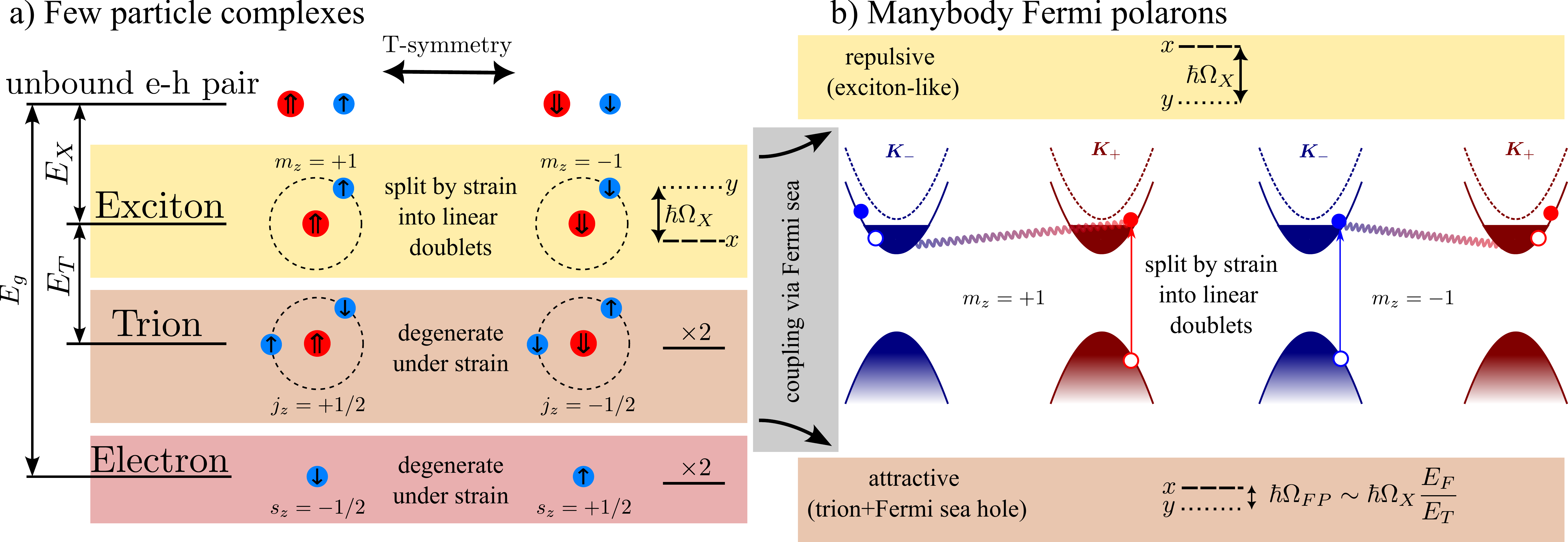}
    \caption{Illustration of the fine structure of Coulomb-correlated states in atomically thin semiconductors. a) Few particle complexes. The states related by the time reversal symmetry are shown. Exciton states with $m_z=\pm 1$ are split by the strain into linearly polarized combinations, Eq.~\eqref{exc:xy}. Electron and trion states are characterized by half-integer spin and degenerate under strain. b) Manybody Fermi polaron approach. The states underlying optical response are repulsive, exciton-like, and attractive, trion-like, Fermi polarons described by superpositions of optically active excitons and electron-hole excitations in the Fermi sea. Both attractive and repulsive polarons have $m_z=\pm 1$ and can be split by the strain. The time reversal symmetry of intervalley Fermi polaron is shown.}
    \label{fig:Tsymmetry}
\end{figure}

The fine structure of quasiparticle energy spectrum and its modification in the presence of strain can be conveniently analyzed from the symmetry standpoint. To highlight important differences in the fine structure of bosonic and fermionic quasiparticles, namely, excitons and trions, illustrated in Fig.~\ref{fig:Tsymmetry}, we start from a simplified model of a monolayer structure with an axial symmetry. While TMDC MLs possess only three-fold rotation axis and are described by the $D_{3h}$ point symmetry group, it is instructive to consider the higher symmetry model which captures all relevant effects. We assume, for simplicity, that the conduction and valence bands are simple and two-fold Kramers degenerate. Corresponding states in doublets related by the time reversal (T) symmetry are denoted for electrons and holes by the effective spin-$1/2$ $z$-component, where $z$ is the ML normal: $s_z=\pm 1/2$ both for electron and hole. According to the general rules~\cite{ll3_eng,edmonds,varshalovich} under time reversal these spin states transform as follows
\begin{equation}
    \label{TimeReversal:electron}
    \mathrm T |s_z\rangle = (-1)^{1/2+s_z} |-s_z\rangle.
\end{equation}
As a result, the state $|+1/2\rangle$ passes to the state $-|-1/2\rangle$, while the state $|-1/2\rangle$ passes to the state $+|-1/2\rangle$. It is noteworthy that due to this rule, there is no time reversal invariant linear combination of the spin-$1/2$ particle states. The same naturally holds true for any fermionic states.

In TMDC MLs Kramers degenerate states are formed in the opposite valleys $\bm K_\pm$, see inset in Fig.~\ref{fig:Tsymmetry}b), and we use the following convention: For electron, $s_z=+1/2$ corresponds to the occupied state in the conduction band in the $\bm K_+$ valley and $s_z=-1/2$ corresponds to the occupied state in the $\bm K_-$ valley. Similarly, for hole $s_z=\pm 1/2$ correspond to the \emph{unoccupied} states in the $\bm K_\pm$ valley. Such convention allows to represent optical selection rules under circularly polarized exciton in the most straightforward way: $\sigma^+$ polarized photon propagating along the $z$-axis induces optical transition in the $\bm K_+$ valley and creates an electron-hole pair with spin-$z$ components of both charge carriers being $+1/2$, while $\sigma^-$ polarized photon induces optical transition in the $\bm K_-$ valley and creates a pair where both electron and hole have $s_z=-1/2$. Hence, the $z$-component of the exciton total angular momentum (or exciton pseudospin) $m_z$, being the sum of the electron, $s_z^{(e)}$, and hole, $s_z^{(h)}$, spin projections, $m_z = s_z^{(e)}+s_z^{(h)}= \pm 1$ for the $\sigma^\pm$ excitation. Naturally, the exciton states $|m_z=\pm 1 \rangle$ are related by the time reversal symmetry, 
\begin{equation}
     \label{TimeReversal:exciton}
\mathrm T|m_z\rangle = |-m_z\rangle.
\end{equation}
Note that in this case a prefactor with the power of $-1$ is absent because the exciton states belong to the multiplet with the total angular momentum $m=1$, thus both $(-1)^{m+1}$ and $(-1)^{m-1}$ are unity.
The combinations of the states
\begin{equation}
    \label{exc:xy}
    |x\rangle = - \frac{|+1\rangle - |-1\rangle}{\sqrt{2}}, \quad 
        |y\rangle = \mathrm i \frac{|+1\rangle + |-1\rangle}{\sqrt{2}},
\end{equation}
are linearly polarized along the $x$ and $y$ axes (we use canonical representation for the angular momentum states). Making use of Eq.~\eqref{TimeReversal:exciton} one can readily see that under time reversal $\mathrm T|x\rangle = -|x\rangle$ and $\mathrm T|y\rangle = |y\rangle$, i.e., these states are time reversal invariant. In the presence of elastic strain, a time reversal invariant perturbation, applied in a such a way that $x$ and $y$ are the main axes of the strain tensor $u_{\alpha\beta}$ ($\alpha$, $\beta$ are Cartesian subscripts), the linearly polarized states $|x\rangle$ and $|y\rangle$ are the eigenstates of the system and the splitting between these states 
\begin{equation}
    \label{Exciton:strain}
    \hbar\Omega_X = \hbar \mathcal B(u_{xx} - u_{yy}),
\end{equation} 
appears, where $\mathcal B$ is the parameter, see Ref.~\cite{PhysRevB.106.125303} for details. For relevant strain values $|\hbar\Omega_X| \lesssim 1$~meV~\cite{PhysRevB.106.125303}.

Let us now turn to the analysis of the trion energy spectrum. We focus on the $X^-$ trion containing two electrons and a hole, the results for the $X^+$ state with two holes and electrons are essentially the same. The $X^-$ trion states can be conveniently described by the $z$-component of the total spin of three constituting charge carriers: two electrons and a hole:
\begin{equation}
    \label{jz:trion}
    j_z= s_z^{{(e_1)}}+s_z^{{(e_2)}} +s{_z^{(h)}}.
\end{equation}
The total spin of two electrons can be either $0$ or $1$.
Since the wave function should be antisymmetric with respect to the permutations of two identical fermions, electrons in our case, the state with the total electron spin $0$ (singlet state) corresponds to the antisymmetric spin part and symmetric orbital part of the wave function. Such so-called \emph{symmetric} trion state~\cite{Courtade:2017a} is bound for any mass ratio of the electron and hole~\cite{Stebe:1989aa,Sergeev:2001aa,PhysRevB.88.045318,PhysRevLett.114.107401,Courtade:2017a}. This is the trion state relevant for the optical absorption experiments: Following the absorption of a circularly polarized photon, a (virtual) exciton with $m_z=\pm 1$ picks out an electron with $s_z=\mp 1/2$ from the ensemble of resident carriers to form a two-electron singlet, see Ref.~\cite{Glazov:2020wf} and references therein. As a result, a trion with $j_z=+1/2$ is formed by the $\sigma^+$ photon and the trion with $j_z=-1/2$ is formed by the $\sigma^-$ photon, see Fig.~\ref{fig:Tsymmetry}a). 
According to the time reversal rules for the fermions, Eq.~\eqref{TimeReversal:electron}, such states cannot be split by a strain, which is time reversal invariant perturbation. Since the ground state of the system, electron, also remains degenerate in the presence of strain, optical transitions in this case have the same energies for both linear polarizations.\footnote{The analysis of irreducible representations for the $D_{3h}$ point group relevant for the negatively and positively charged excitons, Tables I and II of Ref.~\cite{Courtade:2017a}, shows that these conclusions hold in the realistic model of monolayer semiconductor.}

The strain can mix, in principle, these \emph{symmetric} trion states with other, generally, unbound trion states with \emph{antisymmetric} orbital wave function and symmetric spin part of the two-electron wave function. Such a mixing results in the energy shifts of the symmetric trion by $\sim (\hbar\Omega_X)^2/E_T \ll |\hbar\Omega_X|$ with $E_T$ being the trion binding energy and does not result in the splitting of the $j_z =\pm 1/2$ \emph{symmetric} trion doublet.

The analysis above demonstrates that if electrons and trions were localized, e.g., on the static disorder, donor or acceptor impurities~\cite{Semina_2022}, or in moir\'e quantum dots~\cite{Rivera:2018aa} with one electron per localization site, the strain does not result in the splitting of the trion states similarly to the case of trion localization in conventional semiconductor quantum dots~\cite{bayer2002}.

The situation is drastically different if a Fermi sea of free resident electrons is present, Fig.~\ref{fig:Tsymmetry}b). To form a trion exciton can pick up any electron with appropriate spin component leaving behind a Fermi sea hole. The most appropriate picture to describe such situation is that of Fermi polarons, excitons bound to the electron-hole excitations in the Fermi sea~\cite{Rapaport:2001uq,PSSB:PSSB343,suris:correlation,Sidler:2016aa,PhysRevB.95.035417,Glazov:2020wf}. The Fermi polaron quasi-particle is bosonic and described, similarly to the exciton, by the total angular momentum component $m_z = \pm 1$ for optically active states. Indeed, an electron-hole excitation in a Fermi sea has a spin zero\footnote{Unoccupied state in a Fermi sea has the same spin component as the electron since this pair is formed in the same valley. According to the general time reversal rule~\cite{birpikus_eng} the hole spin has an opposite sign.}, so the total angular momentum of the Fermi polaron is the same as of the exciton. Hence, in the presence of strain the proper eigenfunctions of Fermi polarons are expected to take the form of the linearly polarized combinations, Eq.~\eqref{exc:xy}. The fine structure of Fermi polarons is controlled by the fine structure of underlying excitons and the presence of the Fermi sea. Thus, for the repulsive polaron -- the exciton-like quasiparticle -- the strain induced splitting is given at low electron densities by Eq.~\eqref{Exciton:strain}. For the attractive polaron -- the trion-like quasiparticle -- the strain induced splitting is expected to scale linearly both with $\Omega_X$ and the electron Fermi energy $E_F$: \begin{equation}
    \label{splitting:FP}
    \hbar\Omega_{FP} \sim \hbar\Omega_X \frac{ E_F}{E_T},
\end{equation} 
where $E_T$ is a characteristic trion energy. The expression~\eqref{splitting:FP} is confirmed by the microscopic theory presented in the following sections.

\section{Model}\label{sec:model}

\subsection{Hamiltonian and general formalism}\label{sec:Ham}

\begin{figure}[b]
    \centering
\includegraphics[width=0.8\linewidth]{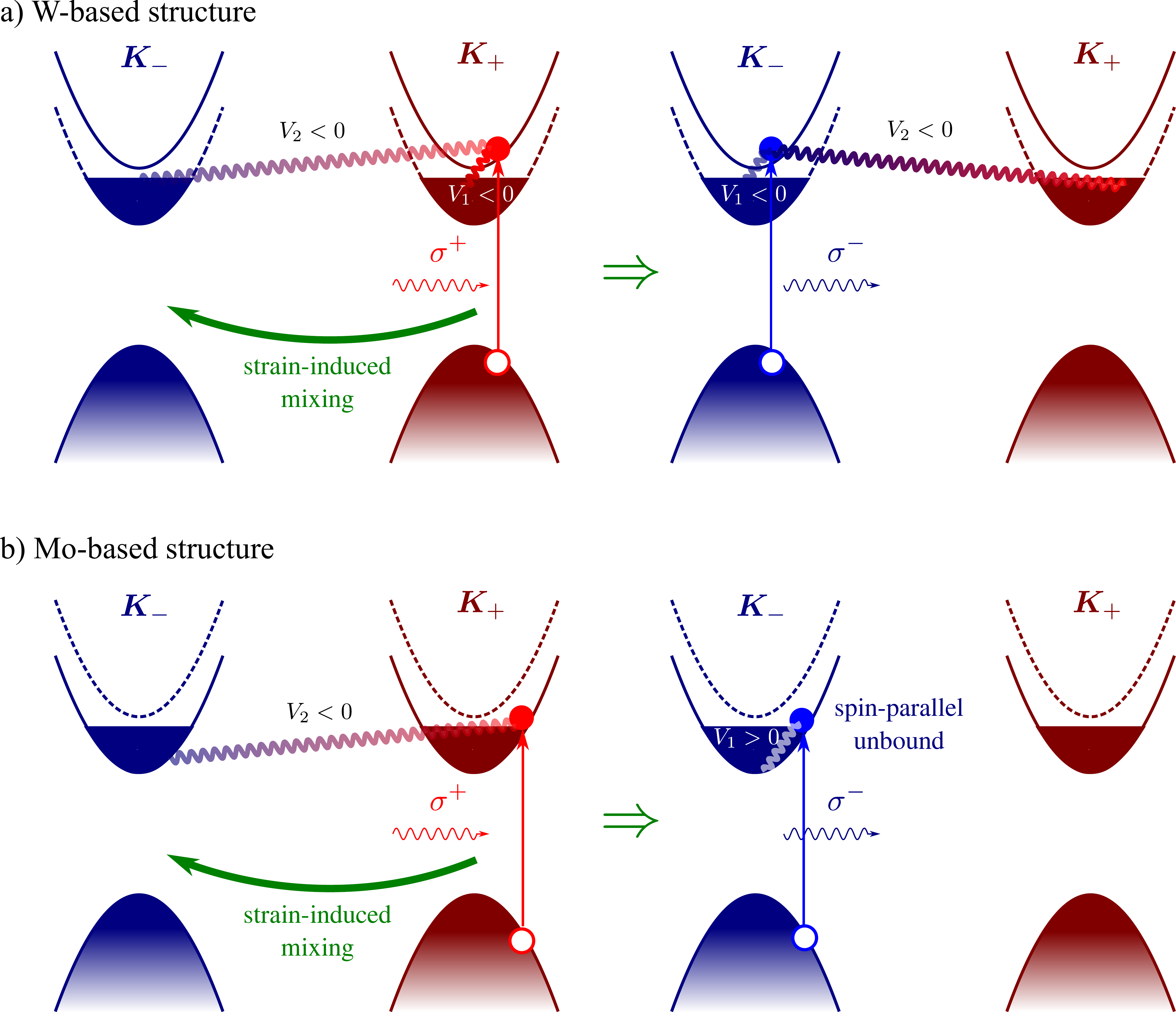}
    \caption{Illustration of the band structure of tungsten-based [panel (a)] and molybdenum-based [panel (b)] monolayers and Fermi polaron formation. Optical transitions active at the normal incidence of light are shown, they involve upper conduction subband in the case of W-based structure and bottom conduction subband in the case of Mo-based structure. Inter- and intra-valley exciton-electron interactions are shown by wavy lines. Green arrows illustrate strain-induced mixing of excitons in opposite valleys.}
    \label{fig:strain:both}
\end{figure}

We start the model description with a case of a tungsten-based monolayer where the spin-orbit splitting in conduction band is such that the transition between the topmost valence band subband and the bottom conduction band subband is forbidden at a normal incidence of light. The optical transitions involve formation of excitons with electrons in the excited conduction subband, Fig.~\ref{fig:strain:both}a). We assume that the electron density is sufficiently low that only bottom conduction subbands are filled with electrons, i.e., the electron Fermi energy $E_F$ reckoned from the botton subband is smaller than the conduction band spin-orbit splitting. We also assume that the Fermi energy is much smaller than the exciton and trion binding energies. This is typical situation in experiments~\cite{PhysRevB.105.075311,PhysRevLett.125.267401,wagner:trions,PhysRevB.105.085302,Robert:2021wc}.

The system's Hamiltonian~$\mathcal{H}$ consists of three  contributions
\begin{equation}
\label{H:tot}
    \mathcal{H} = \mathcal{H}_e + \mathcal{H}_X + \mathcal{H}_{eX},
\end{equation}
where $\mathcal{H}_e$ is the  Hamiltonian of the electrons, $\mathcal{H}_X$ is the Hamiltonian of the excitons and $\mathcal{H}_{eX}$ is the electron-exciton interaction part of the Hamiltonian. 

We write Hamiltonian in the second quantization representation. The system contains two valleys: $\bm K_+$ where the optical transitions are caused by the photons of right circular polarization and $\bm K_-$ where the optical transitions are caused by the photons of left circular polarization, Fig.~\ref{fig:strain:both}. We take into account only one orbital state of excitons, $1s$, since remaining exciton states are relatively far in energy. We denote the creation and annihilation operators for excitons formed by the right circularly polarized light as $R_{\bm k}^\dagger$ and $R_{\bm k}$, we  use the notations $a_{\bm k}^\dagger$, $a_{\bm k}$ for the creation and annihilation operators for electrons in the ground subband in the $\bm K_+$ valley. Accordingly, the creation and annihilation operators~$L_{\bm k}^\dagger$, $L_{\bm k}$ and $b_{\bm k}^\dagger$, $b_{\bm k}$, respectively, refer to the left circularly polarized excitons and the electrons in the bottom conduction subband in the $\bm K_-$ valley. Here, $\bm k$ is the in-plane quasiparticle's wave vector. The Hamiltonian of the non-interacting electrons is
\begin{equation}
\label{H:e}
    \mathcal{H}_e = \sum_{\bm k}\eps_{\bm k}\left(a_{\bm k}^\dagger a_{\bm k} + b_{\bm k}^\dagger b_{\bm k}\right),
\end{equation}
where $\eps_{\bm k} = \h^2k^2/(2M_e)$ is the kinetic energy of the non-interacting electron with the effective mass~$M_e$. Following~\cite{PhysRevB.106.125303}, we consider the Hamiltonian of the excitons in presence of the in-plane strain as
\begin{equation}
\label{H:X}
    \mathcal{H}_X = \sum_{\bm k}\left({\eps_{\bm k}^X\left[R_{\bm k}^\dagger R_{\bm k} + L_{\bm k}^\dagger L_{\bm k}\right]} + \frac{\h\Omega_X}{2}R_{\bm k}^\dagger L_{\bm k} + \frac{\h\Omega_X}{2}L_{\bm k}^\dagger R_{\bm k}\right).
\end{equation}
Here the kinetic energy of the exciton $\eps_{k}^{X} = h^2k^2/(2M_X)$ corresponds to the non-interacting exciton with mass~$M_X$ and wave vector~$\bm k$. We take the energy of exciton with $\bm k = 0$ as a zero energy level. The $\h\Omega_X$ is the strain-induced contribution that mixes right- and left-circularly polarized excitons, Eq.~\eqref{Exciton:strain}. In accordance with the analysis presented above we assume that $x$ and $y$ are the main in-plane axes of the strain tensor and the parameter $\Omega_X$ is real. We recall that strain affects only excitonic states in the Hamiltonian~\eqref{H:X}.


We use the minimum model of the exciton-electron interaction~\cite{suris:correlation,Glazov:2020wf}: We assume that excitons are `rigid' and do not change their orbital state at the scattering by electrons, thus we disregard formation of trions and Fermi polarons associated with excited excitonic states~\cite{PhysRevLett.125.267401}. We use a contact interaction model to describe the exciton-electron coupling, see Refs.~\cite{2019arXiv191204873F,PhysRevB.103.075417} for more elaborate approaches. There are two contributions to the electron-exciton interaction: the intravalley and the intervalley interaction described by the constants $V_1$ and $V_2$, respectively, see wavy lines in Fig.~\ref{fig:strain:both}. Thus, the interaction Hamiltonian is
\begin{eqnarray}
\label{H:eX}
    \mathcal{H}_{eX} & = & V_1\sum_{\bm k, \bm k', \bm p, \bm p'}\delta_{\bm k + \bm p, \bm k' + \bm p'}\left(R_{\bm k'}^\dagger a_{\bm p'}^\dagger R_{\bm k}a_{\bm p} + L_{\bm k'}^\dagger b_{\bm p'}^\dagger L_{\bm k}b_{\bm p}\right) \nonumber \\ & + & V_2\sum_{\bm k, \bm k', \bm p, \bm p'}\delta_{\bm k + \bm p, \bm k' + \bm p'}\left(R_{\bm k'}^\dagger b_{\bm p'}^\dagger R_{\bm k}b_{\bm p} + L_{\bm k'}^\dagger a_{\bm p'}^\dagger L_{\bm k}a_{\bm p}\right),
\end{eqnarray}
where $\delta_{\bm k, \bm p}$ is the Kronecker $\delta$-symbol, which describes momentum conservation at the exciton-electron scattering. The parameters $V_{1,2}$ are related to the trion binding energies, see below.

We consider the case of the zero temperature\footnote{Additional effects related to finite temperatures can be studied following approach of Ref.~\cite{tiene2022crossover}, these are beyond the scope of the present paper.}. The ground state electron Fermi sea wave function is denoted as $|FS\rangle$, it corresponds to the bottom conduction subbands in both $\bm K_+$ and $\bm K_-$ valleys filled up to $E_F$. Optical excitation of the system creates excitons which, owing to the interaction~\eqref{H:eX}, perturb the Fermi sea and create electron-hole pairs transferring an electron from a state below the Fermi energy to the state above $E_F$. In the lowest approximation only one electron-hole pair in the Fermi sea is excited (see Refs.~\cite{PhysRevLett.101.050404,LanLobo} where the role of additional pairs is discussed) and the wave function of the correlated exciton-electron state -- Fermi polaron -- can be recast as~\cite{suris:correlation,PhysRevA.74.063628,PhysRevB.105.075311}
\begin{eqnarray}
\label{FP:wave:gen}
    \left|\Psi_{\bm k}\right\rangle = \varphi^R_{\bm k}R_{\bm k}^\dagger|FS\rangle + \varphi^L_{\bm k}L_{\bm k}^\dagger|FS\rangle + \sum_{\bm p, \bm q}F_{\bm k}^{RR}(\bm p, \bm q)\hat{R}_{\bm k - \bm p + \bm q}^\dagger \hat{a}_{\bm p}^\dagger \hat{a}_{\bm q}|FS\rangle \nonumber \\ + \sum_{\bm p, \bm q}F_{\bm k}^{RL}(\bm p, \bm q)\hat{R}_{\bm k - \bm p + \bm q}^\dagger \hat{b}_{\bm p}^\dagger \hat{b}_{\bm q}|FS\rangle + \nonumber \\
    \sum_{\bm p, \bm q}F_{\bm k}^{LL}(\bm p, \bm q)\hat{L}_{\bm k - \bm p + \bm q}^\dagger \hat{b}_{\bm p}^\dagger \hat{b}_{\bm q}|FS\rangle  + \sum_{\bm p, \bm q}F_{\bm k}^{LR}(\bm p, \bm q)\hat{L}_{\bm k - \bm p + \bm q}^\dagger \hat{a}_{\bm p}^\dagger \hat{a}_{\bm q}|FS\rangle.
\end{eqnarray}
Here $\bm k$ is the Fermi polaron translational motion wave vector,
 $\varphi^{R,L}_{\bm k}$ are the coefficients of wave function of the bare right or left circularly polarized exciton, $F^{RR}_{\bm k}(\bm p, \bm q)$ and $F^{RL}_{\bm k}(\bm p, \bm q)$ are the coefficients describing the intra- and intervalley admixtures of electron-hole pair excitations to the right polarized excitons [$F^{LL}_{\bm k}(\bm p, \bm q)$ and $F^{LR}_{\bm k}(\bm p, \bm q)$ are the same coefficients describing the intra- and intervalley admixtures of the electron-hole pairs to the left polarized excitons]
 with the $\bm p$ and $\bm q$ being the wave vectors of the electron above the Fermi energy and of the unoccupied state in the Fermi sea, respectively. 
 Here and in what follows we use the following convention: in sums the states denoted by $\bm p$, $\bm p'$, etc. are assumed to be above the Fermi energy, i.e., $p,p'\geq k_F$, while the states denoted by $\bm q$, $\bm q'$, etc. are assumed to be below the Fermi energy, $q,q' \leq k_F$ with $k_F = \sqrt{2M_e E_F}/\h$.

The Schr\"odinger equation with the Hamiltonian~\eqref{H:tot}, $\mathcal H|\Psi_{\bm k}\rangle = E_{\bm k} \Psi_{\bm k}$, where $E_{\bm k}$ is the Fermi polaron energy transforms to the system of linear equations for wave function coefficients $\varphi^{R,L}_{\bm k}$, $F^{RR}_{\bm k}(\bm p, \bm q)$, etc. They read for the excitonic part of wave function:
\begin{subequations}
\label{system:full}
\begin{equation}
    \label{full_phiR}
    \varepsilon_{\bm k}^X\varphi_{\bm k}^R + V_1\sum_{\bm p, \bm q}F_{\bm k}^{RR}(\bm p, \bm q) + V_2\sum_{\bm p, \bm q}F_{\bm k}^{RL}(\bm p, \bm q) + {\sum_{\bm q}V_1\varphi_{\bm k}^R + \sum_{\bm q}V_2\varphi_{\bm k}^R} + \frac{\hbar\Omega_X}{2}\varphi_{\bm k}^L = E_{\bm k}\varphi_{\bm k}^R,
\end{equation}
where the mixing of the right- and left- circularly polarized excitons is described by the term $\propto \hbar\Omega_X$ and admixture of the electron-hole pair excitations to the exciton by the terms $\propto V_1, V_2$. The admixture coefficients of the wave function obey
\begin{eqnarray}
    \label{full_FRR}
    \left(\varepsilon_{\bm k - \bm p + \bm q}^X + \varepsilon_{\bm p} - \varepsilon_{\bm q}\right)F_{\bm k}^{RR}(\bm p, \bm q) + V_1\varphi_{\bm k}^R + V_1\sum_{\bm p'}F_{\bm k}^{RR}(\bm p', \bm q) \nonumber \\ +  V_1\sum_{\bm q'}F_{\bm k}^{RR}(\bm p, \bm q') + \frac{\hbar\Omega_X}{2}F_{\bm k}^{LR}(\bm p, \bm q) = E_{\bm k}F_{\bm k}^{RR}(\bm p, \bm q),
\end{eqnarray}
\begin{eqnarray}
    \label{full_FLR}
    \left(\varepsilon_{\bm k - \bm p + \bm q}^X + \varepsilon_{\bm p} - \varepsilon_{\bm q}\right)F_{\bm k}^{LR}(\bm p, \bm q) + V_2\varphi_{\bm k}^L + V_2\sum_{\bm p'}F_{\bm k}^{LR}(\bm p', \bm q) \nonumber \\ + V_2\sum_{\bm q'}F_{\bm k}^{LR}(\bm p, \bm q') + \frac{\hbar\Omega_X}{2}F_{\bm k}^{RR}(\bm p, \bm q) = E_{\bm k}F_{\bm k}^{LR}(\bm p, \bm q)
\end{eqnarray}
\end{subequations}
Equations for $\varphi_{\bm k}^L$, $F_{\bm k}^{LL}(\bm p, \bm q)$, and $F_{\bm k}^{RL}(\bm p, \bm q)$ can be derived from the expressions above interchanging $\{R \leftrightarrow L\}$.  In agreement with the convention formulated above in these equations summation over $\bm p$, $\bm p'$ is carried out under the condition $p, p' \geq k_F$ while over $\bm q$, $\bm q'$ is carried out under the condition $q, q' \leq k_F$. 

Set of Eqs.~\eqref{full_phiR}, \eqref{full_FRR}, and \eqref{full_FLR} together with analogous equations with $\{R \leftrightarrow L\}$ provides energies and wave functions of attractive (trion-like) and repulsive (exciton-like) Fermi polarons. It is convenient to calculate the Green's function of the system by adding source terms $I^R$ ($I^L$) in the left-hand side of Eq.~\eqref{full_phiR} (and of analogous equation for $\varphi^L$). The matrix Green's function $\hat{\mathcal G}_E(\bm k)$ gives the linear response of the system as
\begin{equation}
\label{exciton:phi:greens}
\begin{pmatrix}
\varphi^R_{\bm k}\\
\varphi^L_{\bm k}
\end{pmatrix} = 
\hat{\mathcal G}_E(\bm k) 
\begin{pmatrix}
I^R\\
I^L
\end{pmatrix},
\end{equation}
and allows us to calculate the optical absorption spectra $\propto \Im \hat{\mathcal G}_E(0)$, the poles of the Green's function provide the Fermi polaron eigenenergies. Taking into account that in the presence of the strain with the main axes $x$ and $y$ the eigenpolarizations are linear, Eq.~\eqref{exc:xy}, we pass to corresponding linearly polarized combinations of excitonic states 
\begin{equation}
    \label{linear_basis}
    \varphi^X = {-\frac{\varphi^R-\varphi^L}{\sqrt{2}}}, \quad \varphi^Y = {\mathrm i\frac{\varphi^R+\varphi^L}{\sqrt{2}},}
\end{equation}
and decouple equations for the $X$ and $Y$ polarization. The matrix Green's function becomes diagonal in the linear basis with the components\footnote{We set $\bm k=0$ hereafter since at the normal incidence of light only excitons with $\bm k=0$ are optically active.}
\begin{eqnarray}
    \label{Glinear}
    \mathcal G^{XX}_E = \frac{1}{E + \mathrm i\Gamma {+} \frac{\hbar\Omega_X}{2} - \Sigma(E +\mathrm i\gamma, \hbar\Omega_X)}, \nonumber \\ 
    \mathcal G^{YY}_E = \frac{1}{E + \mathrm i\Gamma {-} \frac{\hbar\Omega_X}{2} - \Sigma(E +\mathrm i\gamma, -\hbar\Omega_X)},
\end{eqnarray}
where $\Gamma$ and $\gamma$ are phenomenological dampings of the exciton and trion states and the self-energy $\Sigma$ accounts for the exciton-electron interaction effects. The latter takes the form (see ~\ref{sec:appendix:der} for derivation)
\begin{equation}
\label{self:energ}
    \Sigma(E, \hbar\Omega) = \sum_{\bm q}\frac{{V_1}}{1 - V_1S(\bm q)} + \sum_{\bm q}\frac{{V_2}}{1 - V_2S(\bm q)} - \sum_{\bm q}\frac{2V_1V_2S_2(\bm q)}{\left[1 - V_1S(\bm q)\right]\left[1 - V_2S(\bm q)\right]}.
\end{equation}
Here we introduced the following notations 
\begin{subequations}
    \label{notations}
\begin{equation}
\label{sumS}
    S(\bm q) = \sum_{p\geq k_F}\zeta(\bm p, \bm q) = -\mathcal{D}\ln{\frac{E_X}{-E - \frac{\hbar^2q^2}{2\mu} + \frac{M_T}{M_X}E_F}},
\end{equation}
\begin{equation}
\label{sumS2}
    S_2(\bm q) = \sum_{p \geq k_F}\zeta^2(\bm p, \bm q)\frac{\hbar\Omega_X}{2} \approx S_2(0) \approx -\mathcal{D}\frac{\hbar\Omega_X}{2E},
\end{equation}
where 
\begin{equation}
\label{zeta}
    \zeta (\bm p, \bm q) = \frac1{E - \eps^X_{\bm q - \bm p} - \eps_{\bm p} + \eps_{\bm q}},
\end{equation}
\end{subequations}
is the Green's function of the non-interacting electron-exciton pair,
$1/\mu = 1/M_e - 1/M_T$, where $M_T = M_X + M_e$ is the trion mass and $\mathcal{D}$ is the exciton-electron reduced density of states, and $E_X$ is the cut-off energy on the order of exciton binding energy. Hereafter it is assumed that $E_F, \hbar\Omega_X \ll E_T$, and we are interested only in the vicinity of the  attractive Fermi polaron (trion) resonance in the spectra, thus $E \approx - E_T$. Note, that dependence of $S_2(\bm q)$ on $\bm q$ and $E_F$ impacts the results only on the next order in small parameters $|\h \Omega_X|/E_T$, $E_F/E_T$ and is not important for the present work. 

The self-energy in Eq.~\eqref{self:energ} has a transparent physical meaning. The first two terms are the self-energies of the intra- and inter- valley Fermi polarons in the absence of the exciton fine structure splitting. In that case exciton interacts independently with electrons in the same and opposite valley. The last term is specific strain-induced term as $S_2(\bm q) \propto \hbar\Omega_X$. It describes the mixing of the Fermi polarons from different valleys due to the exciton splitting. It can be understood as a combination of two processes (which leads to the factor of 2) each of those contains of three steps. The first process is the excitation of intravalley Fermi polaron, changing the valley for the exciton due to the mixing $\propto \hbar\Omega_X$, which changes the type of Fermi polaron to intervalley and formation of the new intervalley polaron. The second process is a reversed first one and consists in the conversion from the intervalley to intravalley polaron. 

The self-energy~\eqref{self:energ} can be explicitly calculated by substituting expansions for $S(\bm q)$, $S_2(\bm q)$ from Eqs.~\eqref{sumS}, \eqref{sumS2} and performing analytical integration. Making use of the relations between the intra- and inter-valley trion binding energies, $E_{T_1}$ and $E_{T_2}$, and the interaction matrix elements~\cite{suris:correlation,Glazov:2020wf}
\begin{equation}
    \label{trion:binding}
    E_{T_{1,2}} = E_X\exp{\left(\frac{1}{\mathcal D V_{1,2}} \right)},
\end{equation}
and introducing the average trion energy $E_T$ and the splitting between the trions $\Delta$ 
\begin{equation}
    \label{trions:E:notations}
E_T = \frac{E_{T_1}+E_{T_2}}{2}, \quad \Delta = |E_{T_1}-E_{T_2}|
\end{equation}
we get
\begin{eqnarray}
\label{Sigma:W:final}
    \Sigma(E, \hbar\Omega_X) = \left[\left(1 {-} \frac{\hbar\Omega_X}{\Delta}\right)\ln\left(1 + \frac{M_X}{M_T}\frac{E_F}{E + E_T - \frac{M_T}{M_X}E_F - \frac{\Delta}{2}}\right) \nonumber \right. \\ \left. + \left(1 {+} \frac{\hbar\Omega_X}{\Delta}\right)\ln\left(1 + \frac{M_X}{M_T}\frac{E_F}{E + E_T - \frac{M_T}{M_X}E_F + \frac{\Delta}{2}}\right)\right]\left(\frac{M_T}{M_X}\right)^2E_T.
\end{eqnarray}
In Eq.~\eqref{Sigma:W:final} we assumed that $E_{T_2}<E_{T_1}$.
Equations~\eqref{Glinear} and \eqref{Sigma:W:final} provide the solution for the fine structure of the attractive  Fermi polaron problem. Its analysis is presented below in Sec.~\ref{sec:results}.

We recall that the self-energy \eqref{Sigma:W:final} is derived for the case of W-based monolayers where optically active exciton can form bound trion state with both electrons in the same valley and in the opposite one: both interaction parameters $V_1$ and $V_2$ are negative. One can readily extend this approach to the case of Mo-based monolayers where only inter-valley trion is bound since optically active exciton is formed from the electrons in the bottom conduction band, Fig.~\ref{fig:strain:both}b). Hence, the two-electron Bloch function in the same valley is symmetric with respect to the electron permutation resulting in the antisymmetric envelope function. Thus, in this case, $V_2<0$, but $V_1>0$. Since exciton-electron scattering amplitude in the same valley has no poles for negative energies~\cite{suris:correlation}, the term describing intravalley Fermi polaron can be safely neglected. Indeed, calculation shows that the pole is present for $E>0$ (i.e., in the exciton $+$ e-h pairs continuum) with the energy $E_r \sim E_X \exp{[1/(\mathcal D V_1)]} \sim E_T(E_X/E_T)^2>0$~\cite{suris:correlation}, i.e., at huge positive energies beyond the validity of our model. The correction resulting from the mixing term $\propto V_1V_2$ is parametrically small, see~\ref{sec:appendix:SigmaMo}. In that case Eqs.~\eqref{Glinear}, and the self-energy can be obtained from Eq.~\eqref{self:energ} with $V_1=0$ with the result [cf. Eq.~\eqref{Sigma:W:final}]
\begin{equation}
\label{Sigma:Mo:final}
    \Sigma(E) = \left(\frac{M_T}{M_X}\right)^2E_T \ln\left(1 + \frac{M_X}{M_T}\frac{E_F}{E + E_T - \frac{M_T}{M_X}E_F}\right).
\end{equation}
For molybdenium-based monolayers the self-energy does not depend on the exciton fine structure splitting. 

Note that the case of positive charged Fermi polarons where the system contains resident holes rather then electrons is the same as Mo-based case, Eq.~\eqref{Sigma:Mo:final} with $E_T$ being corresponding positively charged trion $X^+$ binding energy and $E_F$ being the hole Fermi energy. It is due to the considerable spin-orbit splitting of the valence band. Hence, only the intervalley interaction is important.

\subsection{Trion pole approximation\label{trion:pole}}

It is worth noting that Eqs.~\eqref{Glinear} for the matrix Green's function and self-energies Eqs.~\eqref{self:energ} can be derived directly using the diagrammatic approach~\cite{suris:correlation,Glazov:2020wf,PhysRevX.9.041019,PhysRevB.105.075311}. To that end it is convenient to introduce the bare exciton matrix Green's function which accounts for the exciton fine structure
\begin{equation}
\label{G:E:matrix}
\hat{G}_{E}(\bm k) = \frac{1}{E - \varepsilon_{\bm k}^X - \frac{\hbar\Omega_X}{2} \hat{\sigma}_x},
\end{equation}
where $\hat{\sigma}_x$ is the pseudospin Pauli matrix and $2\times 2$ unit matrix is omitted. In this section we use the representation of the pseudospin Pauli matrices such that the eigenstate of $\hat{\sigma}_x$ with the eigenvalue $-1$ corresponds to the $x$-polarized exciton. Correspondingly, the matrix $\hat{S}(\bm q)$, the exciton-electron loop, takes in the leading order in $\h\Omega_X$ the form
\[
\hat{S}(\bm q) = \sum_{p\geq k_F} \hat{G}_{E-\varepsilon_{\bm p} + \varepsilon_{\bm q}}(\bm q - \bm p) = S(\bm q) + \hat{\sigma}_x S_2(\bm q),
\]
where $S(\bm q)$ and $S_2(\bm q)$ are given by Eqs.~\eqref{notations}. The scattering matrix $\hat{T}$ is found from the standard equation which treats exactly exciton interaction with a single electron (we omit argument $\bm q$ for brevity)
\begin{equation}
\label{T:matrix:gen}
\hat{T} = \hat{V} + \hat{V}\hat{S}\hat{T}, \quad \hat{V} = \begin{pmatrix}
    V_1 & 0 \\
    0 & V_2
\end{pmatrix}.
\end{equation}
Solving Eq.~\eqref{T:matrix:gen} we obtain in the first order in $\h\Omega_X/E_T$
\begin{equation}
    \label{T:matrix:sol}
\hat{T} = \frac{1}{1-\hat{V}S}\hat{V} + \frac{1}{1-\hat{V}S} \hat{V} S_2 \hat{\sigma}_x \hat{V}\frac{1}{1-\hat{V}S}.
\end{equation}
This solution yields the exciton Green's function
\begin{equation}
    \label{GG}
    \hat{\mathcal G}_E(0) = \frac{1}{E - \varepsilon_{\bm k}^X - \frac{\hbar\Omega_X}{2} \hat{\sigma}_x - \sum_{q\leq k_F} \left(\hat{T} +  \{1 \leftrightarrow 2\}\right)}.
\end{equation}
Note that the additional term in the denominator with the replacement $V_1$ by $V_2$ and vice versa takes into account that exciton interacts with two Fermi seas: in the same and other valley. 
In the linearly polarized basis, the Green's functions and self-energies take the form of Eqs.~\eqref{Glinear} and \eqref{self:energ}.

Such an approach allows us to formulate, following Refs.~\cite{Glazov:2020wf,PhysRevB.105.075311} (see also Ref.~\cite{imamoglu2020excitonpolarons}), a simplified -- \emph{trion pole} -- approximation which captures basic physics of the effect. To that end the scattering matrix in the vicinity of $E\approx -E_T$ is approximated by a pole describing the bound exciton-electron (trion) state and the summation over $\bm q$ in the self-energy is replaced by the multiplication by electron density $N_e$. For the tungsten-based structures it yields the self-energy in the form
\begin{subequations}
\label{Sigma:pole}
\begin{equation}
\label{Sigma:W:pole}
    \Sigma(E, \hbar\Omega_X) = \alpha E_F E_T \left[
   \frac{1 {-} \h\Omega_X/\Delta}{E + E_T - \beta E_F - \frac{\Delta}{2}} + \frac{1 {+} \h\Omega_X/\Delta}{E + E_T - \beta E_F + \frac{\Delta}{2}}\right],
\end{equation}
where the parameters $\alpha$ and $\beta$ are determined from the condition that the Fermi polaron energies from Eq.~\eqref{Sigma:W:pole} are equal to those from the general expression for the self-energy~\eqref{Sigma:W:final} at $E_F \ll \Delta, E_T$:
\begin{equation}
    \alpha = \frac{(M_X/M_T)^3}{4\sinh^2\left[\frac{1}{2}\left(\frac{M_X}{M_T}\right)^2\right]}, \qquad \beta = \alpha+\frac{M_T}{M_X} - \frac{M_X/M_T}{1 - \exp\left[-\left(\frac{M_X}{M_T}\right)^2\right]}.
\end{equation}
For molybdenium-based systems the self-energy takes the single-pole form:
  \begin{equation}
\label{Sigma:Mo:pole}
    \Sigma(E) = 
   \frac{\alpha E_F E_T}{E + E_T - \beta E_F}.
\end{equation}  
\end{subequations}
These expressions clearly show that the exciton self-energy has a resonant feature at the trion energy (with the shift $\beta E_F$ related to its modification by the presence of the electron gas) and that the strength of the exciton-trion coupling is $\propto \sqrt{E_F E_T}$~\cite{Glazov:2020wf,imamoglu2020excitonpolarons}.

This model allows us to obtain simple closed-form relations for the attractive Fermi polaron energies. For the W-based systems solution of  the equation $E=\Sigma(E,\hbar\Omega_X)$ for the poles of the Green's function yields quadruplet of the states stemming from the intra and intervalley trions:
\begin{subequations}
\label{trion:pole:poles}
\begin{equation}
\label{polarons:W:poles}
    E_{FP} = -E_T + (\beta - \alpha)E_F \pm \sqrt{\frac{\Delta^2}{4} + (\alpha E_F)^2 \pm \alpha E_F\hbar\Omega_X},
\end{equation}
where the contributions on the order of $\hbar\Omega_X/E_T$ are neglected. For the Mo-based systems we have for the energies of the attractive polaron doublet
\begin{equation}
\label{polarons:Mo:poles}
    E_{FP} = -E_T + (\beta - \alpha)E_F \pm \alpha E_F \frac{\hbar\Omega_X}{2E_T}.
\end{equation}
\end{subequations}

To conclude the model description we emphasise the main  assumptions behind our model: First, it is assumed that the electron Fermi energy $E_F$ is much smaller than the trion binding energy $E_T$: $E_F \ll E_T$. The latter is, in its turn, much smaller than the exciton binding energy. Second, the strain induced exciton fine structure splitting $\h\Omega_X$ is assumed to be much smaller than trion binding energy $|\h\Omega_X| \ll E_T$. Third, the self-energies derived above either in full form, Eq.~\eqref{Sigma:W:final},~\eqref{Sigma:Mo:final}, or in the simplified trion-pole approximation, Eqs.~\eqref{Sigma:pole}, are valid for $|E-E_T| \ll E_T$, i.e., in the vicinity of the attractive Fermi polaron. Note that in the vicinity of the exciton resonance ($E\approx 0$) the main effects of the Fermi sea are the shift and the broadening of exciton resonance, as well as reduction of its oscillator strength~\cite{suris:correlation,PhysRevB.95.035417,wagner:trions}. The corrections to the exciton fine structure splitting due to the presence of electron Fermi sea are unimportant for this work.

\section{Results and discussion}\label{sec:results}

\subsection{Attractive Fermi polaron fine structure}

The Green's function found above allow us to calculate the eigenstates of the system and optical response. Let us start with analyzing the energy spectrum fine structure of attractive Fermi polarons. We first consider a more general case of W-based structure with the band structure and optical transitions illustrated in Fig.~\ref{fig:strain:both}a).

\begin{figure}
    \centering
    \includegraphics[width=0.8\linewidth]{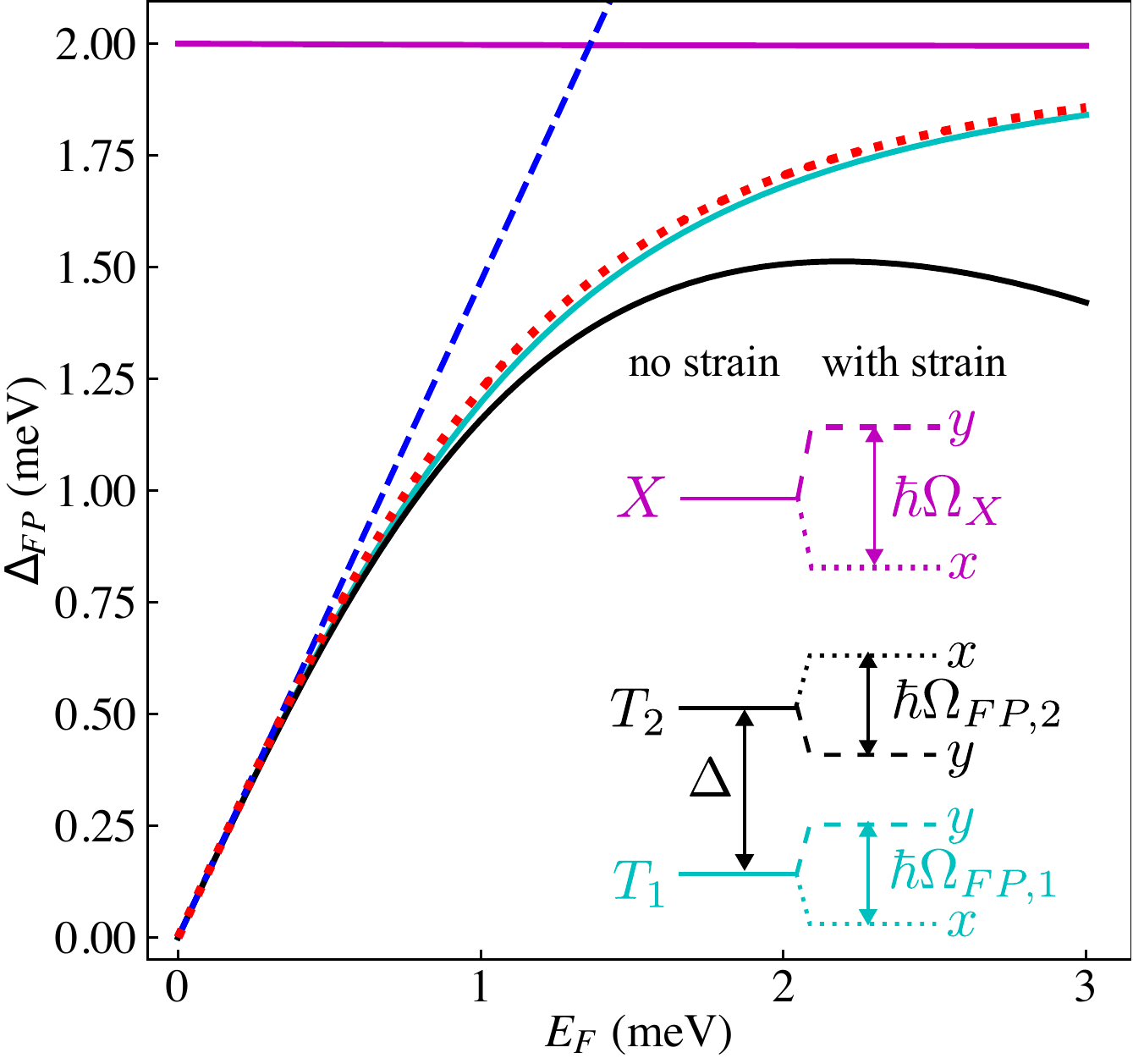}
    \caption{Fermi polaron energy splitting. Solid cyan and black lines show, respectively, the splitting of the first, $T_1$, and second, $T_2$, attractive Fermi polaron states calculated numerically from Eq.~\eqref{FullE0} for W-based monolayer. Solid magenta line shows the exciton (repulsive Fermi polaron) splitting $\hbar\Omega_X$. Blue dashed line shows the linear asymptotics for small Fermi energy, Eq.~\eqref{DeltaFP}. Red dotted line shows analytical result for the attractive Fermi polaron splitting within the trion pole approximation, Eq.~\eqref{polarons:W:poles}, that tends to the exciton splitting at sufficiently large $E_F$. Inset shows schematics of the attractive and repulsive Fermi polaron fine structure in the absence and presence of the strain. The order of energies of inter- and intravalley trions is selected in agreement with Ref.~\cite{Robert:2021wc}. The parameters of calculations are as follows: $E_T = 20$meV, $\Delta = 4$meV, $\h\Omega_X = 2$meV, effective masses $M_e = M_h$.}
    \label{fig:DeltaFP}
\end{figure}

The Fermi polarons energies correspond to the poles of~Eq.~\eqref{Glinear}. Equation $E=\Sigma(E,\h\Omega_X)$ in the vicinity of the attractive Fermi polaron (trion) resonance, $E \approx - E_T$, reduces to
\begin{eqnarray}
    \label{FullE0}
    \left(\frac{M_X}{M_T}\right)^2 + \left(1 {\mp} \frac{\hbar\Omega_X}{\Delta}\right)\ln\left(1 + \frac{M_X}{M_T}\frac{E_F}{{E} + E_T - \frac{M_T}{M_X}E_F - \frac{\Delta}{2}}\right) \nonumber \\ + \left(1 {\pm} \frac{\hbar\Omega_X}{\Delta}\right)\ln\left(1 + \frac{M_X}{M_T}\frac{E_F}{{E} + E_T - \frac{M_T}{M_X}E_F + \frac{\Delta}{2}}\right) = 0,
\end{eqnarray}
where the top and bottom signs correspond to $x$- and $y$-linearly polarized states, see inset in Fig.~\ref{fig:DeltaFP}. There are four roots of Eq.~\eqref{FullE0}. Calculated splittings of the inter- and intra-valley states are plotted in Fig.~\ref{fig:DeltaFP} as functions of the Fermi energy for the fixed exciton fine structure splitting $\h\Omega_X$. Naturally, for $E_F\to 0$ the attractive Fermi polaron states are not split, since the trion states are degenerate, see Sec.~\ref{sec:symmetry}. The splitting increases with the increase in the electron density due to the enhancement of the exciton-trion coupling. Interestingly, the order of the fine structure split levels is different for two attractive Fermi polarons: for the lower in energy (stemming from the $T_1$ trion) it coincides with the order of excitonic sublevels, while for the upper one (stemming from the $T_2$ trion) it is opposite. 

In the linear regime where the Fermi energy is sufficiently small, $E_F \ll \Delta$, the attractive Fermi polaron energies can be recast as
\begin{equation}
    E= -E_T \pm \frac{\Delta}{2} + \left(\frac{M_T}{M_X} - \frac{M_X}{M_T}\frac1{1 - \exp\left[-\frac{\Delta}{\Delta {\mp} \hbar\Omega_X}\left(\frac{M_X}{M_T}\right)^2\right]}\right)E_F.
\end{equation}
First two terms in this expression describe the Fermi polaron energies in the absence of strain, the last term $\propto E_F$ describes the strain-induced splitting of intra- and intervalley Fermi polarons.
In this regime the magnitude splitting of each doublet is the same and reads 
\begin{equation}
\label{DeltaFP:0}
    \hbar\Omega_{FP} = \frac{M_X}{M_T}\frac{\sinh\left[\frac{\hbar\Omega_X\Delta}{\Delta^2-(\hbar\Omega_X)^2}\left(\frac{M_X}{M_T}\right)^2\right]}{2\sinh\left[\frac{\Delta}{2(\Delta - \hbar\Omega_X)}\left(\frac{M_X}{M_T}\right)^2\right]\sinh\left[\frac{\Delta}{2(\Delta + \hbar\Omega_X)}\left(\frac{M_X}{M_T}\right)^2\right]}E_F.
\end{equation}
In the realistic situation the strain-induced splitting of excitons is small compared to $\Delta$: $\hbar\Omega_X \ll \Delta$, as a result, Eq.~\eqref{DeltaFP:0} reduces to
\begin{equation}
    \label{DeltaFP}
    \hbar\Omega_{FP} = \frac{\left(M_X/M_T\right)^3}{2\sinh^2\left[\frac1{2}\left(\frac{M_X}{M_T}\right)^2\right]}\frac{\hbar\Omega_X}{\Delta}E_F.
\end{equation}
One can see that $\hbar\Omega_{FP}$ is linear both in the Fermi energy and exciton splitting and scales as $\h\Omega_X E_F/\Delta$. 
Asymptotic Eq.~\eqref{DeltaFP} also directly follows from the simplified trion pole approximation, Eqs.~\eqref{polarons:W:poles}, in the limit of $\hbar\Omega_X\to 0$.
The analysis shows that this contribution $\hbar\Omega_{FP}\propto \hbar\Omega_X/\Delta$ stems from the $\Omega_X$-linear terms in the exciton self-energy described by the last term in Eq.~\eqref{self:energ}. For both intra- and intervalley Fermi polarons the $\Delta$ in denominator comes from the difference of the poles energies. 

It is instructive to consider also a limit of $E_{T_1} = E_{T_2}$, i.e., the case where both inter- and intravalley trions are degenerate.\footnote{Experimentally, the bare splitting $E_{T_1} - E_{T_2}$ can be controlled, e.g., by the isotropic strain $u_{xx} = u_{yy}$.} The most straightforward way to obtain the attractive polaron energies is to use Eq.~\eqref{polarons:W:poles} but take into account that for $\hbar\Omega_X \gtrsim \Delta$ one has two redefine $\Delta$ as $\Delta = \sqrt{(E_{T_1}-E_{T_2})^2+\hbar\Omega_X^2}$. Hence, we obtain two doublets with linear in $E_F$ splitting:
\begin{equation}
\label{polarons:W:poles:0}
    E_{FP} = -E_T + (\beta - \alpha)E_F \pm \alpha E_F \pm \frac{\hbar\Omega_X}{2}.
\end{equation}
In this particular case the attractive Fermi polarons at $E_F=0$ form two degenerate doublets spit by $\hbar\Omega_X$. With increase in $E_F$ each of the doublets splits into linearly polarized components with the splitting given by $2\alpha E_F$.

In Mo-based structures, similarly to conventional quantum wells, only one trion state is bound, see Fig.~\ref{fig:strain:both}b) and discussion above. The analysis of the self-energy Eq.~\eqref{Sigma:Mo:final} shows that in the limit $E_F \ll E_T$
\begin{equation}
\label{DeltaFP:Mo}
\h\Omega_{FP} = {\frac{\left(M_X/M_T\right)^3}{4\sinh^2\left[\frac1{2}\left(\frac{M_X}{M_T}\right)^2\right]}} \frac{\hbar\Omega_X}{E_T}E_F.
\end{equation}
Generally, in this case the splitting is smaller than for W-based structures by the factor $\Delta/E_T$. This is because in this situation the strain mixes the bound and unbound polaron states with the energy distance $\sim E_T$. Again, Eq.~\eqref{DeltaFP:Mo} is in agreement with the trion-pole approximation, Eq.~\eqref{polarons:Mo:poles}.

The linear in $\h\Omega_X$ contributions to the attractive Fermi polaron anisotropic splittings  can be readily derived using the first order perturbation theory. To that end we start with the Fermi polaron wavefunctions~\eqref{FP:wave:gen} calculated for $\h\Omega_X \equiv 0$, i.e., where the right- and left- circularly polarized excitons are decoupled. Then we calculate the fine matrix elements of perturbation responsible for the exciton anisotropic splitting $\hat U =  (\h\Omega_x/2) \sum_{\bm k} (R^\dag_{\bm k} L_{\bm k} + L^\dag_{\bm k} R_{\bm k})$, Eq.~\eqref{H:X}. In the case of Mo-based monolayers the only non-zero contribution to the matrix element stemms from the excitonic components of the wavefunction~\eqref{FP:wave:gen}:
\[
\langle \Psi_{\bm k}^L |U| \Psi_{\bm k}^R\rangle = |\varphi_{\bm k}|^2 \frac{\hbar\Omega_X}{2}.
\]
For the polaron with zero wavevector $|\varphi_{\bm 0}|^2 = \alpha E_F/E_T$ gives the ratio of the attractive polaron to exciton oscillator strength yielding Eq.~\eqref{DeltaFP:Mo}. For the W-based monolayers the main contribution to the matrix element comes from the admixtures in the wavefunction~\eqref{FP:wave:gen} of the electron-hole pair excitations within the same and opposite valley. In that case one recovers Eq.~\eqref{DeltaFP}.

\subsection{Polarization-dependent optical spectra}

\begin{figure}[t]
    \centering
    \includegraphics[width=0.8\linewidth]{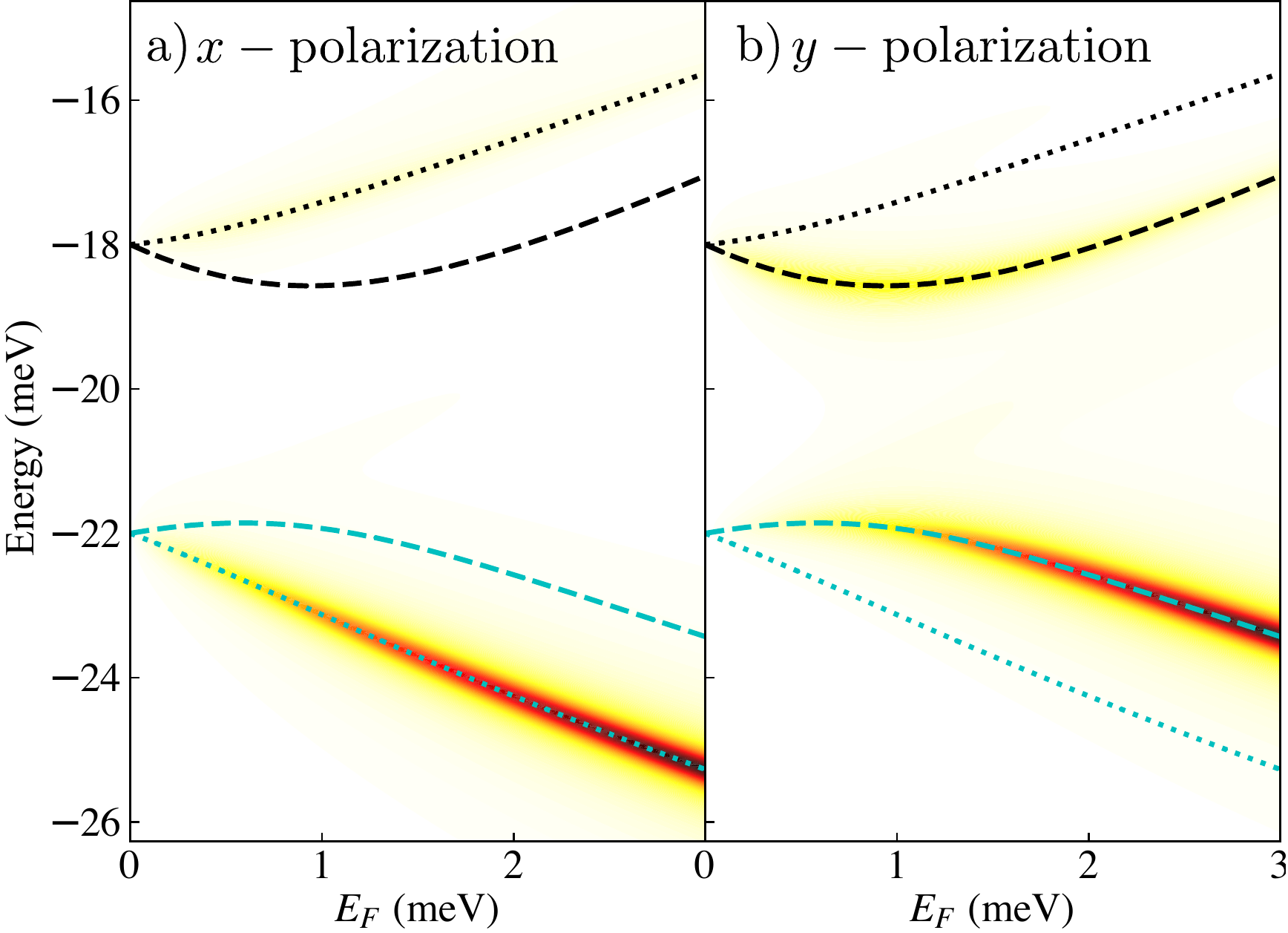}
    \caption{Absorption spectra a) for $x$-polarized light and b) for $y$-polarized light calculated after Eqs.~\eqref{Abs}, \eqref{Glinear}, and \eqref{Sigma:W:final} for W-based monolayer as function of Fermi energy in the spectral range of attractive Fermi polarons. False color scale shows the absorbance $\mathcal A$. The cyan and black lines show energies of the first and second attractive Fermi polaron doublets calculated from Eq.~\eqref{FullE0}. Dotted lines show the $x$-polarized sublevels, while dotted lines show the $y$-polarized sublevels, see inset in Fig.~\ref{fig:DeltaFP}. The parameters are the same as in Fig.~\ref{fig:DeltaFP}.}
    \label{fig:FPabs}
\end{figure}

We now turn to calculations of optical spectra and plot in Fig.~\ref{fig:FPabs} absorbance of the monolayer in the trion (attractive Fermi polaron) spectral range for $x$- and $y$-linear polarizations
\begin{equation}
\label{Abs}
\mathcal A_X \propto - \Im \{\mathcal G_E^{XX}\}, \quad \mathcal A_Y \propto - \Im \{\mathcal G_E^{YY}\},
\end{equation}
with the exciton Green's function is given by Eq.~\eqref{Glinear}. The color shows the intensity of absorption. Lines overlayed on the plot show the positions of the poles of the Green's functions calculated using Eq.~\eqref{FullE0}. Figure~\ref{fig:FPabs} shows that the increase in the electron Fermi energy results both in increase of the oscillator strength of the attractive Fermi polarons and, for not too high $E_F$, of the splitting of the Fermi polarons in the linearly polarized components. In agreement with the symmetry analysis the spectra in a given linear polarization, $x$ or $y$, contains only two lines that are polarized accordingly. The order of states, as discussed above, is different for two attractive Fermi polarons. Our calculation shows that the spectra calculated using the full self-energy, Eq.~\eqref{Sigma:W:final}, and that in the trion pole approximation, Eq.~\eqref{Sigma:W:pole}, are almost identical.

\begin{figure}[ht]
    \centering
    \includegraphics[width=0.8\linewidth]{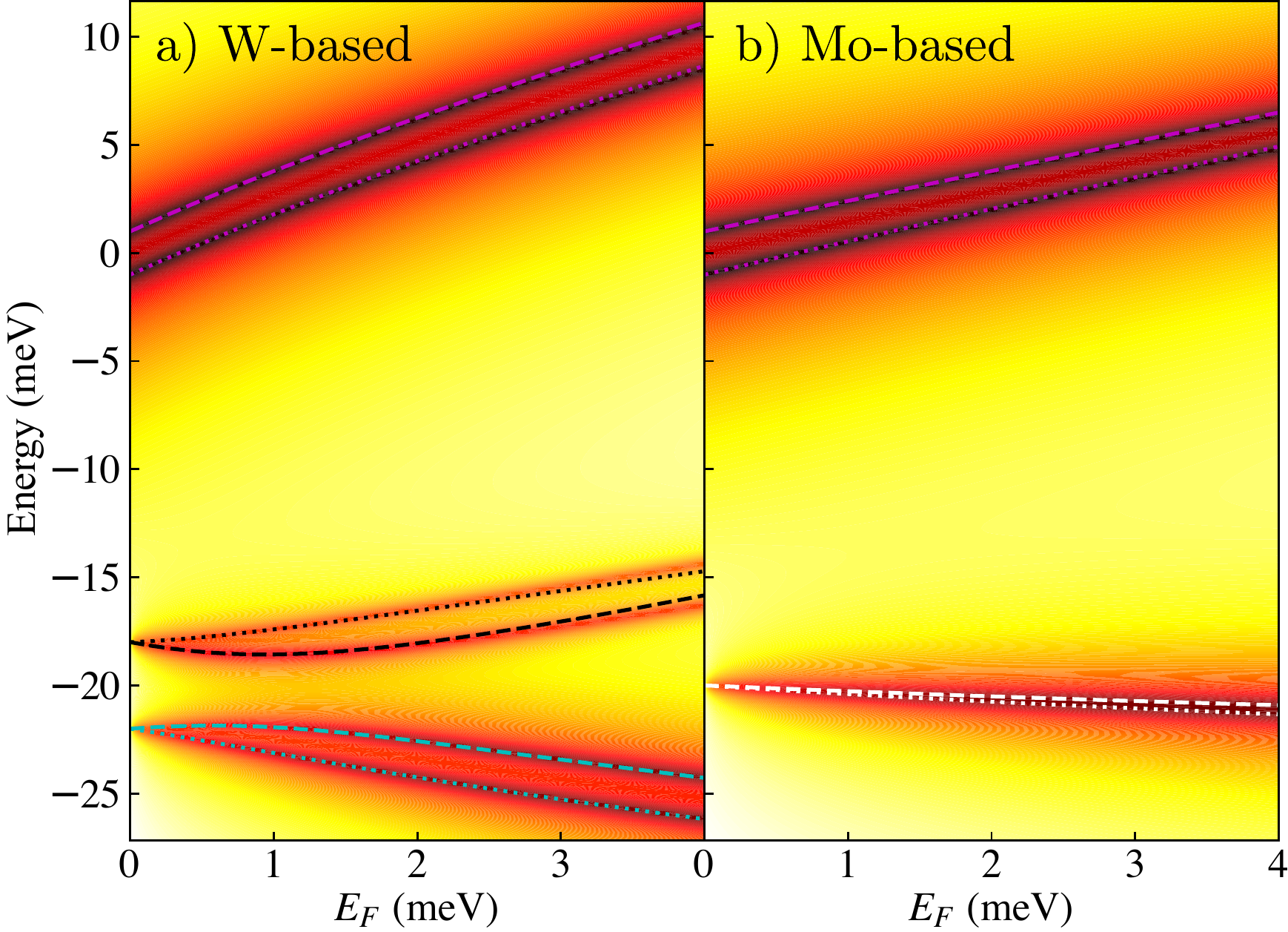}
    \caption{Absorption spectra a) for W-based and b) for Mo-based  structures in circularly polarized light for wide energy range covering both repulsive and attractive Fermi polarons calculated within the trion-pole approximation, Eqs.~\eqref{Abs}, \eqref{Glinear}, and \eqref{Sigma:pole}, as function of Fermi energy. False color scale shows the logarithm of absorbance $\ln\mathcal A$. Magenta lines show the repulsive polaron splitting, while black and cyan lines in a) show the splitting two attractive polarons for W-based monolayer. White dotted and dashed lines in b) corresponds to the $x$- and $y$-polarized attractive Fermi polarons, respectively, in Mo-based monolayer. The parameters are the same as in Fig.~\ref{fig:DeltaFP}; for Mo-based monolayer the trion binding energy $E_T=20$~meV}.
    \label{fig:FPabsAll}
\end{figure}

Figure~\ref{fig:FPabsAll} shows the absorption spectra in circular polarization (note that these spectra are the same in $\sigma^+$ and $\sigma^-$ polarization) for the W-based structure [panel a)] and Mo-based structure [panel b)]. Naturally, the absorption in circular polarization is determined by all states of Fermi polarons, both $x$- and $y$-polarized. More precisely, it is a linear combination of imaginary parts of $\mathcal{G}_E^{XX}$ and $\mathcal{G}_E^{YY}$ in agreement with Eq.~\eqref{linear_basis}. Hence, for the tungsten-based monolayer one can see six lines: two stemming from the exciton radiative doublet and four from the intra- and intervalley trions, while for the molybdenum-based monolayer there are four lines: two from the exciton and two from the trion. In this figure for illustrative purposes we used the wide range of energies covering both exciton and trion resonances and employed the trion pole approximation for the self-energy, Eqs.~\eqref{Sigma:pole}. In agreement with the analytical theory described above, Eqs.~\eqref{DeltaFP} and \eqref{DeltaFP:Mo}, the attractive Fermi polaron splitting is sufficiently larger for W-based system as compared to the Mo-based one because $\Delta\ll E_T$.

In the presence of anisotropic strain, $u_{xx} \ne u_{yy}$, the optical selection rules change and transitions in both $\bm K_+$ and $\bm K_-$ valley become elliptically polarized with the orientation of the ellipse being linked to the main axes of the strain tensor~\cite{PhysRevB.106.125303}. Thus, overall intensities of absorption in $x$- and $y$-linear polarization in Fig.~\ref{fig:FPabs} can differ. Estimates of induced linear polarization $P_l$ in Ref.~\cite{PhysRevB.106.125303} show that $P_l \propto (u_{xx} - u_{yy})$. This effect is disregarded here for simplicity.

\subsection{General remarks and outlook}

We have demonstrated above that attractive Fermi polarons, as integer spin quasiparticles, are subject to the fine structure splitting in the presence of strain. The calculation presented above demonstrates that the splitting of Fermi polarons originates from the excitonic fine structure splitting, Eq.~\eqref{Exciton:strain}. As shown in Ref.~\cite{PhysRevB.106.125303}, the parameter $\mathcal B$ governing excitonic splitting $\Omega_X$ has two contributions: due to the short- and long-range electron-hole exchange interaction. The latter results from the coupling of the electron-hole pair with the induced electromagnetic field~\cite{glazov2014exciton,prazdnichnykh2020control}, and its calculation for attractive Fermi polarons should be done similarly to the calculation for excitons: (i) The long-range contribution should be excluded from $\hbar\Omega_X$ in Eq.~\eqref{Exciton:strain}, and (ii) The long-range splitting of the attractive polarons can then be calculated from Maxwells equations taking into account polaron interaction with induced electromagnetic field. It yields $\hbar\Omega_{FP} \propto P_l \Gamma_{0,FP} \propto P_l\Gamma_{0,X} E_F/E_T$, where $P_l$ is the strain-induced linear polarization of optical transitions due to band mixing~\cite{PhysRevB.106.125303}, $\Gamma_{0,FP}$ and $\Gamma_{0,X}$ are the radiative decay rates of attractive polaron and exciton, respectively.

Let us now briefly address the attractive Fermi polaron fine structure in bilayer systems. In that case typically two types of Coulomb-correlated complexes are present, namely, intra- and interlayer ones. For the intralayer complexes the situation is the same as above: The selection rules are controlled by the monolayer transitions and exciton interaction with the Fermi sea of resident carriers either in the same or in the other layer can be described by the Hamiltonian~\eqref{H:eX}. The analysis shows that the fine structure of the interlayer complexes is also quite similar. The main difference in this case results from the fact that depending on the stacking and particular atomic registry both transitions between spin-like and spin-unlike bands become possible at the normal incidence of radiation~\cite{Yu_2018,Forg:2019aa}. Still, the energy gaps between the bottom and top conduction subbands are significant which allows one to consider corresponding Fermi polarons independently and use the formalism developed above. Qualitatively, the fine structure of the energy spectrum for the Fermi polarons will be the same, quantitatively, however, the splittings are expected to be smaller due to weaker overlap of the electron and hole wave functions and, accordingly, smaller value of $\mathcal B$ in Eq.~\eqref{Exciton:strain} and smaller exciton fine structure splitting $\hbar\Omega_X$. Interestingly, the presence of moir\'e potentials does not only change the exciton and trion emission energy resulting in the formation of superlattice-like or quantum-dot like states, but may also result in variation of the electron density across the sample. The latter results in variation of $\hbar\Omega_{FP}$ in the heterobilayer plane owing to the coordinate dependence of the Fermi energy, see Eqs.~\eqref{DeltaFP} and \eqref{DeltaFP:Mo}.

Specific situation may arise in heterobilayers where electrons and trions are trapped by the potential minima~\cite{Baek:2021aa,PhysRevX.11.031033,Campbell:2022aa}. In the case of $N_e=1$ resident electron per localization site the optical transitions are controlled by the trion formation [electron to trion transition, Fig.~\ref{fig:Tsymmetry}a)] and the fine structure splitting of charged exciton should be absent. If $N_e=2$ resident electrons are present at the localization site then their ground state has a zero spin and the corresponding optical transition can demonstrate the fine structure splitting. Such parity effect depending on the electron number $N_e$ can persist for higher occupancies provided the exchange interaction between the localized electrons is large enough. 

\section{Conclusion}\label{sec:concl}

We have shown that attractive Fermi polarons in two-dimensional semiconductors, by contrast to the trions, demonstrate the fine structure splitting caused by an anisotropic strain in the monolayer plane.  The splitting is inherited from that of the neutral exciton (or repulsive polaron) and scales linearly both with the exciton splitting and electrons or holes Fermi energy. The splittings have been calculated using the ansatz solution of the Schr\"odinger equation for interacting electrons and excitons in the model of the short-range interaction and in the Green's function approach. We have also calculated optical absorption spectra demonstrating linear polarization anisotropy induced by the strain for attractive Fermi polarons. Analytical results for the strain induced splitting of Fermi polarons are obtained both for the W- and Mo-based structures. Our predictions are in line with recent observations of the fine structure splitting in strained MoS$_2$ monolayers deposited on nanomembranes where the anisotropic splitting of the charged exciton peaks in the presence of strain has been observed~\cite{10.1088/2053-1583/ac7c21}.

Similar effects can be also observed in conventional semiconductor quantum wells with weak disorder. For strongly localized charge carriers and trions the fine structure splitting is absent.

Studies of the momentum dependent longitudinal-transverse splitting for Fermi polarons, valley entanglement of Fermi polarons by linearly polarized light, as well as strain-induced Fermi polaron spin dynamics are interesting directions for further research.

\section*{Acknowledgements}

The authors are grateful to Alexey Chernikov and Paulina Plochocka for valuable discussions. Z.A.I. is grateful to the BASIS foundation. This work was supported by RSF project No. 23-12-00142.

\section*{Bibliography}
\bibliographystyle{unsrt}
\bibliography{TFS.bib}

\appendix

\section{Derivation of self-energies}\label{sec:appendix:der}

We will look for the Green's function for the linearly polarised excitation. The Shr\"odinger equation with the source of $x$-polarised light~$I^X$ can be written by taking appropriate linear combinations of Eqs.~\eqref{system:full}
\begin{equation}
    \eps^X_{\bm k}\varphi^X_{\bm k} + V_1\sum_{\bm p, \bm q}F_{\bm k}^{1X}(\bm p, \bm q) + V_2\sum_{\bm p, \bm q}F_{\bm k}^{2X}(\bm p, \bm q) + {\sum_{\bm q}(V_1+V_2)\varphi^X_{\bm k} -} \frac{\hbar\Omega_X}{2}\varphi^X_{\bm k} = E_{\bm k}\varphi^X_{\bm k} - I^X,
\end{equation}
\begin{eqnarray}
    (\eps_{\bm k - \bm p + \bm q}^X + \eps_{\bm p} - \eps_{\bm q})F^{1X}_{\bm k}(\bm p, \bm q) + V_1\varphi_{\bm k}^X + V_1\sum_{\bm p'}F^{1X}_{\bm k}(\bm p', \bm q) \nonumber \\ + V_1\sum_{\bm q'}F^{1X}_{\bm k}(\bm p, \bm q') - \frac{\hbar\Omega_X}{2}F_{\bm k}^{2X}(\bm p, \bm q) = E_{\bm k}F_{\bm k}^{1X}(\bm p, \bm q),
\end{eqnarray}
\begin{eqnarray}
    (\eps_{\bm k - \bm p + \bm q}^X + \eps_{\bm p} - \eps_{\bm q})F^{2X}_{\bm k}(\bm p, \bm q) + V_2\varphi_{\bm k}^X + V_2\sum_{\bm p'}F^{2X}_{\bm k}(\bm p', \bm q) \nonumber \\ + V_2\sum_{\bm q'}F^{2X}_{\bm k}(\bm p, \bm q') - \frac{\hbar\Omega_X}{2}F_{\bm k}^{1X}(\bm p, \bm q) = E_{\bm k}F_{\bm k}^{2X}(\bm p, \bm q),
\end{eqnarray}
where we introduced the linear combinations of wave function coefficients
\begin{eqnarray}
    F_{\bm k}^{1X}(\bm p, \bm q) = {-}\frac{F^{RR}_{\bm k}(\bm p, \bm q) {-} F^{LL}_{\bm k}(\bm p, \bm q)}{\sqrt{2}}, \\ F_{\bm k}^{2X}(\bm p, \bm q) = {-}\frac{F^{RL}_{\bm k}(\bm p, \bm q) {-} F^{LR}_{\bm k}(\bm p, \bm q)}{\sqrt{2}}.
\end{eqnarray}
The similar system of equations holds for $y$-polarized polarons, with the replacement $X \leftrightarrow Y$, where
\begin{eqnarray}
    F_{\bm k}^{1Y}(\bm p, \bm q) = {\mathrm i}\frac{F^{RR}_{\bm k}(\bm p, \bm q) {+} F^{LL}_{\bm k}(\bm p, \bm q)}{\sqrt{2}}, \\ F_{\bm k}^{2Y}(\bm p, \bm q) = {\mathrm i}\frac{F^{RL}_{\bm k}(\bm p, \bm q) {+} F^{LR}_{\bm k}(\bm p, \bm q)}{\sqrt{2}}
\end{eqnarray}
and change of sign of $\Omega_X$.

To solve these equations we assume, that $\bm k = 0$ and the fine structure splittings are small compared to the trion energy: $$\hbar \Omega_X, |E_{T1} - E_{T2}|, E_F \ll E_{T1,2} \ll E_X.$$
We can express $\varphi^X_{\bm k}$ through the sums of the wave function coefficients ($E\equiv E_{\bm k=0}$)
\begin{equation}
    \varphi^X = \frac{V_1\mathcal{F}^{1X}_{p, q} + V_2\mathcal{F}^{2X}_{p, q}}{E {+} \frac{\hbar\Omega_X}{2}{-\sum_{\bm q}(V_1+V_2)}} + \frac{I^X}{0 {+} \frac{\hbar\Omega_X}{2}{-\sum_{\bm q}(V_1+V_2)}},
\end{equation}
where we introduced
\begin{equation}
    \mathcal{F}_{p}(\bm q) = \sum_{\bm p}F(\bm p, \bm q), \quad \mathcal{F}_{q}(\bm p) = \sum_{\bm q}F(\bm p, \bm q), \quad \mathcal{F}_{p, q} = \sum_{\bm p, \bm q}F(\bm p, \bm q).
\end{equation}

The pair of equations for $F^{1,2X}(\bm p, \bm q)$ with assumptions above transforms into the system of equations
\begin{equation}
    \zeta^{-1}(\bm p, \bm q)F^{1X}(\bm p, \bm q) {+} \frac{\hbar\Omega_X}{2}F^{2X}(\bm p, \bm q) = V_1\left(\mathcal{F}^{1X}_p(\bm q) + \varphi^X\right), \quad \{1 \leftrightarrow 2\},
\end{equation}
where $\zeta(\bm p, \bm q)$ is defined in Eq.~\eqref{zeta}. Taking into account that for attractive Fermi polaron $E_0 \approx -E_T$ we neglect the second order term $\zeta^{2}(\bm p, \bm q)(\hbar\Omega_X)^2 \lesssim (\hbar\Omega_X/E_0)^2 \ll 1$ and express $F^{1,2X}(\bm p, \bm q)$ as
\begin{eqnarray}
    F^{1X}(\bm p, \bm q) = \zeta(\bm p, \bm q)V_1\left(\mathcal F^{1X}_p(\bm q) + \varphi^X\right) \nonumber \\ {-} \zeta(\bm p, \bm q)^2\frac{\hbar\Omega_X}{2}V_2\left(\mathcal{F}^{2X}_p(\bm q) + \varphi^X\right), \quad \{1 \leftrightarrow 2\}.
\end{eqnarray}
After the summation over $\bm p$ we get the system of linear equations for~$\mathcal{F}_p^{1,2X}(\bm p, \bm q)$
\begin{eqnarray}
    \left[1 - V_1S(\bm q)\right]\mathcal F^{1X}_p(\bm q) {+} V_2S_2(\bm q)\mathcal{F}^{2X}_p(\bm q) \nonumber \\ = \left[V_1S(\bm q) {-} V_2S_2(\bm q)\right]\varphi^X, \quad \{1 \leftrightarrow 2\},
\end{eqnarray}
where the sums $S(\bm q)$ and $S_2(\bm q)$ are given by Eqs.~\eqref{sumS} and \eqref{sumS2}. Neglecting terms $\propto \left(\hbar\Omega_X\right)^2 \ll 1$ we express the $\mathcal{F}_p^{1,2X}(\bm q)$
\begin{equation}
    \mathcal{F}_p^{1X}(\bm q) = \frac{\left[1 - V_2S(\bm q)\right]V_1S(\bm q) {-} V_2S_2(\bm q)}{\left[1 - V_1S(\bm q)\right]\left[1 - V_2S(\bm q)\right]}\varphi^X, \quad \{1 \leftrightarrow 2\}.
\end{equation}

Finally, we get self-consistent equation for $\varphi^X$
\begin{eqnarray}
    \varphi^X = \frac{V_1\mathcal{F}^{1X}_{p, q} + V_2\mathcal{F}^{2X}_{p, q}}{E {+} \frac{\hbar\Omega_X}{2}{-\sum_{\bm q}(V_1+V_2)}S(\bm q)} + \frac{I^X}{E{+} \frac{\hbar\Omega_X}{2}{-\sum_{\bm q}(V_1+V_2)S(\bm q)}}.
\end{eqnarray}
The excitonic amplitude in the Fermi polaron state with linear source $I^X$ is
\begin{equation}
    \varphi^X = \frac{I^X}{E {+} \frac{\hbar\Omega_X}{2} - \Sigma(E_0, \hbar\Omega_X)},
\end{equation}
in agreement with Eq.~\eqref{Glinear}, where we introduced self-energy $\Sigma(E_0, \hbar\Omega_X)$, Eq.~\eqref{self:energ}.

\section{Self-energy for molybdenum-based layers}\label{sec:appendix:SigmaMo}


In the case of Mo-based semiconductor (or $p$-type monolayer) where $V_1>0$ and $V_2<0$ the exciton self-energy can be also calculated from Eq.~\eqref{self:energ}. Expressing integrals via the trion binding energy $$E_T \equiv E_{T_2} = E_X\exp{\left(\frac{1}{\mathcal D V_2}\right)}$$ and unbound state resonance energy $$E_r = E_X \exp{[1/(\mathcal D V_1)]} \sim E_T(E_X/E_T)^2 \gg E_X, E_T,$$ we obtain the following estimates for the three terms in self-energy at $E\approx - E_T$:
\begin{subequations}
\begin{eqnarray}
    \sum_{\bm q} \frac{{V_1}}{1 - V_1S(\bm q)} \sim \frac{E_F}{\ln\left(\frac{E_r}{E_T}\right)} \to 0, \label{S:first}\\
    \sum_{\bm q} \frac{{V_2}}{1 - V_2S(\bm q)} = \left(\frac{M_T}{M_X}\right)^2E_T \ln\left(1 + \frac{M_X}{M_T}\frac{E_F}{E + E_T - \frac{M_T}{M_X}E_F}\right), \label{S:second} \\
    \sum_{\bm q} \frac{2V_1V_2S_2(\bm q)}{\left[1 - V_1S(\bm q)\right]\left[1 - V_2S(\bm q)\right]} = \nonumber \\ \left(\frac{M_T}{M_X}\right)^2\frac{\hbar\Omega_X}{\ln\left(\frac{E_r}{E_T}\right)} \ln\left(1 + \frac{M_X}{M_T}\frac{E_F}{E + E_T - \frac{M_T}{M_X}E_F}\right)\to 0. \label{S:third}
\end{eqnarray}
\end{subequations}
One can readily see that the first term is non-resonant. The second~\eqref{S:second} and third~\eqref{S:third} contributions have a feature at $E \approx -E_T$. However, the third contribution has a large logarithm $\ln{(E_r/E_T)}$ in the denominator. As a result, it is negligible compared to the  term $\hbar\Omega_X / 2$ from the exciton splitting. As a result, only the second term, Eq.~\eqref{S:second} remains. Hence, to calculate the self-energy in this case $V_1$ can be set to zero.
\end{document}